\documentclass[12pt]{iopart}
\usepackage{graphicx}
\usepackage{color}
\usepackage{iopams}

\newcommand{\dd}  { {\textrm d}}

\begin{document}

%\title[Particle Ratios at High-$p_T$ at LHC energies]{LAST CALL FOR PREDICTIONS FOR LHC}{Particle Ratios at High-$p_T$ at LHC energies}
\title{Particle Ratios at High $p_T$ at LHC Energies}

%%%%%%%%%%%%%%%%%%%%%%%%%%%%%%%%%%%%%%%%%%%%%%%%%%%%%%%%%%%
\author{G G Barnaf\"oldi, P L\'evai}
\ead{bgergely@rmki.kfki.hu}
\address{MTA KFKI RMKI 
%Research Institute for Particle and Nuclear Physics, 
P.O. Box 49, Budapest 1525, Hungary}

\author{B A Cole}
\address{Nevis Laboratory, Columbia University, New York, NY, USA}

\author{G Fai}
\address{Department of Physics, Kent State University, Kent, OH, USA}

\author{G Papp}
\address{Department of Theoretical Physics, ELTE
%E\"otv\"os University,
P\'azm\'any P. 1/A, Budapest 1117, Hungary}

%%%%%%%%%%%%%%%%%%%%%%%%%%%%%%%%%%%%%%%%%%%%%%%%%%%%%%%%%%%%%%%%%%%%%%%%%%
\begin{abstract}
Hadron production has been calculated in a pQCD improved parton model for
 $pp$, $dA$ and heavy ion collisions. We applied KKP and AKK fragmentation
functions. %Hadron ratios -- measurable by the ALICE experiment -- have
%been investigated.  
Our jet fragmentation study shows, that hadron ratios
at high $p_T$ depend on quark contribution mostly and less on the gluonic 
one. This finding can be seen in jet-energy loss calculations, also. We 
display the suppression pattern on different hadron ratios in $PbPb$ 
collisions at LHC energies. 
\end{abstract}
%%%%%%%%%%%%%%%%%%%%%%%%%%%%%%%%%%%%%%%%%%%%%%%%%%%%%%%%%%%%%%%%%%%%%%%%%%
%Uncomment for PACS numbers title message
%\pacs{00.00, 20.00, 42.10}
% Keywords required only for MST, PB, PMB, PM, JOA, JOB? 
%\vspace{2pc}
%\noindent{\it Keywords}: Article preparation, IOP journals
% Uncomment for Submitted to journal title message
%\submitto{\JPA}
% Comment out if separate title page not required
%\maketitle
%\section{Particle Ratios}
%%%%%%%%%%%%%%%%%%%%%%%%%%%%%%%%%%%%%%%%%%%%%%%%%%%%%%%%%%%%%%%%%%%%%%%%%%
%%%%%%%%%%%%%%%%%%%%%%%%%%%%%%%%%%%%%%%%%%%%%%%%%%%%%%%%%%%%%%%%%%%%%%%%%%
%%%%%%%%%%%%%%%%%%%%%%%%%%%%%%%%%%%%%%%%%%%%%%%%%%%%%%%%%%%%%%%%%%%%%%%%%%

The precision of pQCD based parton model calculations was enhanced during 
the last decade. The calculated spectra allow to make predictions not 
only for the hadron yields, but for sensitive particle ratios and nuclear 
modifications. For the calculation of particle ratios new fragmentation 
functions are needed not only for the most produced light mesons, but for 
protons also. From the experimental point of view one requires identified 
particle spectra by RHIC and LHC. Especially the ALICE detector has a 
unique capability to measure identified particles at highest transverse 
momenta via \v{C}herenkov detectors. The $\pi^{\pm}/K^{\pm}$ and 
$K^{\pm}/p(\bar{p})$ ratios can be measured up to $3$ GeV/c and $5$ GeV/c 
respectively.     

\begin{figure}
\begin{center}
\includegraphics[width=7.5truecm,height=8.0truecm]{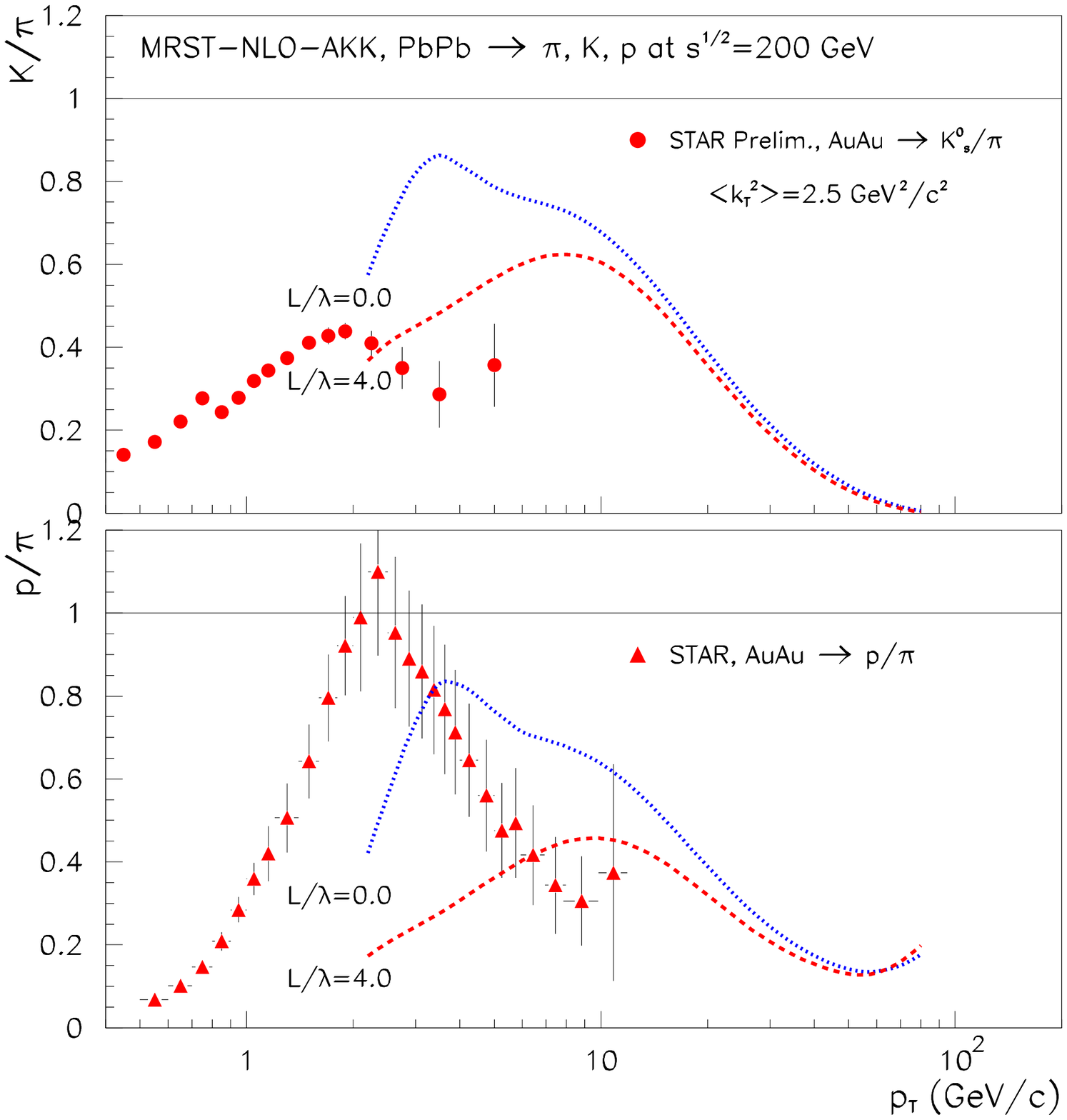}
\includegraphics[width=7.5truecm,height=8.0truecm]{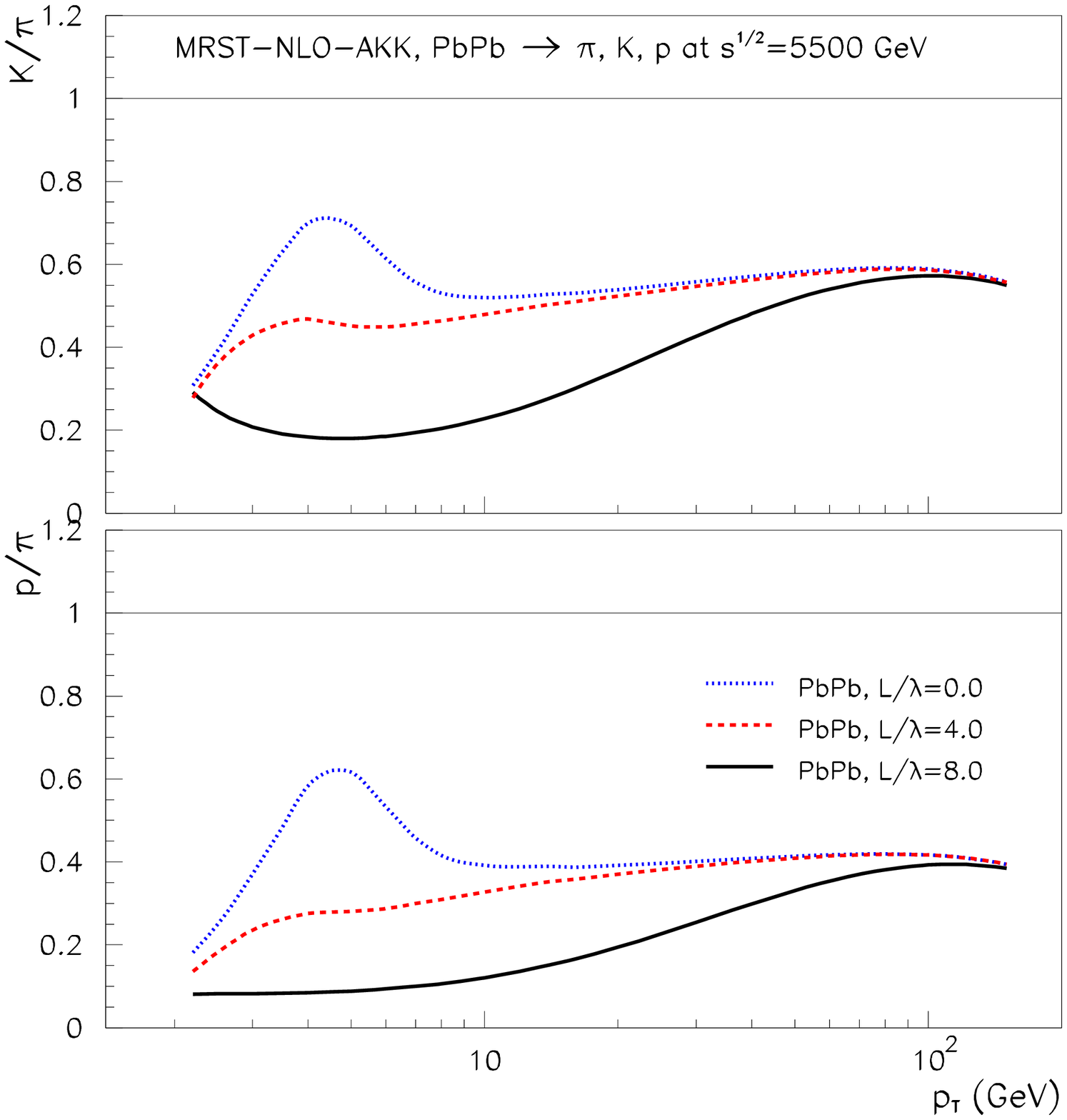}
\end{center}
\vspace*{-1.0truecm}
\caption{Calculated charge-averaged $K/\pi$ and $p/\pi$ ratios in  
$AA$ collisions at RHIC and LHC energies. RHIC curves are compared to 
STAR\cite{STAR_data,STAR_data2} data at $\sqrt{s}=$ 200 $A$GeV.}
\label{fig1}
\end{figure}

Here we calculate hadron ratios in our next-to-leading order pQCD 
improved parton model based on Ref.\cite{bgg_qm06} %~\cite{Yi02} 
with intrinsic transverse momenta, determined by the expected c.m. 
energy evolution along the lines of Ref.\cite{bgg_qm06}. The presented ratios 
are based on $\pi$, $K$ and $p$ spectra which were calculated by AKK 
fragmentation functions~\cite{AKK1}. First we compare calculated particle ratios to the 
data of the STAR collaboration measured in $AuAu$ collisions at 
$\sqrt{s}= 200$ $A$GeV RHIC energy \cite{STAR_data,STAR_data2}. Predictions 
for high-$p_T$ hadron ratios at RHIC and at LHC energies in most central 
($0-10\%$) $PbPb$ collisions are also shown in Fig. \ref{fig1}.   

\newpage

On the {\sl left panel} of Fig. \ref{fig1}, particle ratios are compared 
to $AuAu$ 
collisions at $\sqrt{s}=200$ $A$GeV STAR $K/\pi$ ({\sl dots}) and $p/\pi$ 
({\sl triangles}) data. The agreement between the RHIC 
data and the calculations at RHIC energy can be considered acceptable 
at $p_T \gtrsim 5$ 
GeV/c, with an opacity of $L/\lambda=4$. However, at lower momenta, where 
pQCD is no longer reliable, the ratios differ from the calculated curves.  
 
The {\sl right panel} shows calculations for $PbPb$ collisions for 
$\sqrt{s}= 5.5$ $A$TeV energy. Using a simple 
$\dd N / \dd y \sim 1500-3000$ estimation, we expect a 
$L/\lambda \approx 8$ opacity in most central $PbPb$ collisions. 
For comparison, we plotted the $L/\lambda = 0$ and $4$ values 
also. The lower- and intermediate-$p_T$ variation of the hadron ratios 
arise from the different strengths of the jet quenching for quark and gluon 
contributions\cite{KaonJet}. Due to the quark dominated fragmentation, the 
difference disappears at high-$p_T$ in the ratios.

%%%%%%%%%%%%%%%%%%%%%%%%%%

%%%%%%%%%%%%%%%%%%%%%%%%%%%%%%%%%%%%%%%%%%%%%%%%%%%%%%%%%%%%%%%%%%%%%%%%%%
%\section*{Acknowledgments}
%%%%%%%%%%%%%%%%%%%%%%%%%%%%%%%%%%%%%%%%%%%%%%%%%%%%%%%%%%%%%%%%%%%%%%%%%%
Acknowledgments: Special thanks to Prof. John J. Portman for computer
time at Kent State University. Our work was supported in part by
Hungarian OTKA T047050 and NK62044, by the U.S. Department
of Energy under grant U.S. DE-FG02-86ER40251, and jointly by the U.S.
and Hungary under MTA-NSF-OTKA OISE-0435701.

%%%%%%%%%%%%%%%%%%%%%%%%%%%%%%%%%%%%%%%%%%%%%%%%%%%%%%%%%%%%%%%%%%%%%%%%%%

%%%%%%%%%%%%%%%%%%%%%%%%%%%%%%%%%%%%%%%%%%%%%%%%%%%%%%%%%%%%%%%%%%%%%%%%%%
%%%%%%%%%%%%%%%%%%%%%%%%%%%%%%%%%%%%%%%%%%%%%%%%%%%%%%%%%%%%%%%%%%%%%%%%%%

\section*{References}
\begin{thebibliography}{13}

\bibitem{bgg_qm06}
G.G. Barnaf\"oldi, P. L\'evai, G. Papp, G. Fai and B.A. Cole, 
arXiv:hep-ph/0703059 (2007).
%% CITATION = HEP-PH/0703059;%%

\bibitem{AKK1}
S. Albino, B.A. Kniehl and G. Kramer,
\NP {\bf B725} 181 (2005); {\sl ibid.} {\bf B734} 50 (2006).
%{\tt \verb!http://www.desy.de/~simon/AKK2005FF.html!}
%% CITATION = NPHA,B725,181;%%

\bibitem{STAR_data}
STAR preliminary data, presented by Ming Yao on SQM'07 conference

\bibitem{STAR_data2}
B.I. Adelev {\it et al.} [STAR],
\PRL {\bf 97} 152301 (2006)
%% arXiv:nucl-ex/0607033
%% CITATION = PRLA,97,152301;%%

\bibitem{KaonJet}
P. L\'evai, G. Papp, G. Fai and M. Gyulassy,
{\it Acta. Phys. Hung.} {\bf A27} 459 (2006).
%% CITATION = APHUE,A27,459;%%

\end{thebibliography}
\end{document}